\documentclass[3p]{elsarticle}

\usepackage{lineno,hyperref}
\usepackage{amsmath,amssymb}
\usepackage{subcaption}
\modulolinenumbers[1]

\journal{Journal of Computational Physics}

\begin{document}

\begin{frontmatter}

\title{Eigenstructure-preserving scheme for a hyperbolic system}


\author[roku]{Takashi Shiroto\corref{mycorrespondingauthor}}
\ead{shiroto.takashi@qst.go.jp}
\cortext[mycorrespondingauthor]{Corresponding author}

\author[roku]{Akinobu Matsuyama}
\author[naka,roku]{Nobuyuki Aiba}

\address[roku]{Rokkasho Fusion Institute, National Institutes for Quantum and Radiological Science and Technology,
2-166 Oaza-Obuchi-Aza-Omotedate, Rokkasho, Aomori 039-3212, Japan}
\address[naka]{Naka Fusion Institute, National Institutes for Quantum and Radiological Science and Technology,
801-1 Mukoyama, Naka, Ibaraki 311-0193, Japan}

\begin{abstract}
\end{abstract}


\end{frontmatter}


\section{Introduction}\label{sec:1}
Numerical methods for hyperbolic systems have been discussed for decades
because the inviscid fluid dynamics and ideal magnetohydrodynamics (MHD) are described as hyperbolic systems.
One of the most important milestones in this field is an invention of the upwind method.
The upwind methods are based on the fact that coefficient matrices in the quasi-linear form
have a set of linearly independent eigenvectors and corresponding eigenvalues;
the matrices can be diagonalized and the system can be split into nonlinear wave equations \cite{Toro}.
Such a discussion of the eigenstructure is reflected to the Roe solver \cite{Roe1981}, for example.
The upwind schemes have enabled us to capture a shock wave as a weak solution in the inviscid fluid simulation.
However, the present computational fluid dynamics (CFD) community encounters various numerical instabilities.
One of the examples is the so-called ``carbuncle phenomenon'' \cite{Robinet2000}.
This has been recognized as a numerical instability of shock capturing schemes;
a stable bow-shock can be numerically unstable especially when the Mach number---
a non-dimensional quantity of the ratio between flow speed and sound speed---
is much greater than unity.
What is worrisome is there are some experiments that the bow-shock can be physically unstable
at high Mach number \cite{Ohnishi2015,Kikuchi2017}.
Therefore, it is difficult to judge whether a bow-shock instability in CFD is physical or numerical one.

In MHD simulations of fusion plasmas, the numerical instabilities have been discussed in a different context.
When the linearized MHD equations are discretized, unphysical instabilities are induced by truncation errors;
the numerical phenomenon is called as ``spectral pollution'' \cite{Gruber}.
The spectral pollution is a numerical phenomenon that the eigenvalues of the linearized MHD equations
converge to unphysical values, and the unphysical mode can interfere the other physical modes.
To eliminate the spectral pollution, a hybrid element method was proposed
in which a function space of the finite-element method is designed consistently \cite{Grimm1976,Gruber1981}.
However, no numerical scheme for the nonlinear MHD equations has been proposed to eliminate the spectral pollution.
The nonlinear MHD equation must be transformed into a quasi-linear formulation to analyze the eigenstructure,
but the conservation laws are usually violated in the non-conservative formulation.
Therefore, it is difficult to maintain the eigenstructure and conservation laws simultaneously.

In this article, an eigenstructure-preserving scheme is proposed for a one-dimensional compressible Euler equation, as a first step.
We employ the quadratic conservative scheme \cite{Morinishi2010,Halpern2018,Shiroto2019} to reconcile the
non-conservative and conservative formulations, and to ensure that the discrete system has an unpolluted
eigenstructure and the conservation laws are strictly preserved.
The rest of this article is composed as follows.
The mathematical basis of the eigenstructure-preserving scheme is explained in Sec.~\ref{sec:2}.
A brief numerical experiment is performed to show that no spectral pollution is observed with the proposed scheme in Sec.~\ref{sec:3}.
Section~\ref{sec:4} states the conclusions of this article.

\section{Mathematical basis}\label{sec:2}
Here we discuss the following compressible Euler equation in one-dimension:

\begin{align}
\frac{\partial \mathbf{Q}}{\partial t}+\frac{\partial \mathbf{F}}{\partial x}=\mathbf{0},\label{eq:2.1}\\
\mathbf{Q}=\begin{bmatrix}
\rho \\ \rho u \\ \rho H/\gamma +(\gamma-1)\rho u^2/2\gamma
\end{bmatrix},\ 
\mathbf{F}=\begin{bmatrix}
\rho u\\ (\gamma-1)\rho H/\gamma+(\gamma+1)\rho u^2/2\gamma \\ \rho Hu
\end{bmatrix},\label{eq:2.2}
\end{align}
where $\rho$ is the mass density, $u$ is the fluid velocity, and $H$ is the specific enthalpy.
The pressure and sound speed are defined by the ideal equation of state as follows
while they have already been eliminated from the system:

\begin{align}
p=\frac{\gamma-1}{\gamma}\left(\rho H-\frac12\rho u^2\right),\ 
c_\mathrm{s}=\sqrt{(\gamma-1)\left(H-\frac12u^2\right)}.\notag
\end{align}

Before the derivation of the eigenstructure-preserving scheme, we summarize the strategy of this investigation.
The following two points must be maintained to preserve the eigenstructure and conservation laws simultaneously:

\begin{enumerate}
   \item The conservation laws are preserved with discretizing the quasi-linear form.
   \item The Jacobian matrix is composed consistently so that the characteristic equation is factorized.
\end{enumerate}
It is well known that the skew-symmetric scheme can preserve the conservation laws
with discretizing the non-conservative form \cite{Morinishi2010}.
The gonverning equations in conservative form Eq.~(\ref{eq:2.1}) can be transformed into the
following form by introducing a parameter vector $\mathbf{W}$:

\begin{align}
\frac{\partial \mathbf{Q}}{\partial \mathbf{W}}\frac{\partial \mathbf{W}}{\partial t}+
\frac{\partial \mathbf{F}}{\partial \mathbf{W}}\frac{\partial \mathbf{W}}{\partial x}=\mathbf{0}.\label{eq:2.4}
\end{align}
If the parameter vector $\mathbf{W}$ can describe both $\mathbf{Q}$ and $\mathbf{F}$ in its quadratic form,
the skew-symmetric scheme is applicable to maintain the conservation laws.
Actually, such a parameter vector has already been proposed in the classical work of Roe \cite{Roe1981}:

\begin{align}
\mathbf{W}=\begin{bmatrix}
\sqrt{\rho}\\ \sqrt{\rho}u \\ \sqrt{\rho}H
\end{bmatrix},\ 
\frac{\partial \mathbf{Q}}{\partial \mathbf{W}}=
\begin{bmatrix}
2\sqrt{\rho} & 0 & 0\\
\sqrt{\rho}u & \sqrt{\rho} & 0\\
\dfrac{1}{\gamma}\sqrt{\rho}H & \dfrac{\gamma-1}{\gamma}\sqrt{\rho}u & \dfrac{1}{\gamma}\sqrt{\rho}
\end{bmatrix},\ 
\frac{\partial \mathbf{F}}{\partial \mathbf{W}}=
\begin{bmatrix}
\sqrt{\rho}u & \sqrt{\rho} & 0\\
\dfrac{\gamma-1}{\gamma}\sqrt{\rho}H & \dfrac{\gamma+1}{\gamma}\sqrt{\rho}u & \dfrac{\gamma-1}{\gamma}\sqrt{\rho}\\
0 & \sqrt{\rho}H & \sqrt{\rho}u
\end{bmatrix}.\label{eq:2.3}
\end{align}
It is natural that all components of $\partial \mathbf{Q}/\partial \mathbf{W}$ and $\partial \mathbf{F}/\partial \mathbf{W}$
are expressed only with those of $\mathbf{W}$.
This formulation is similar to a numerical method proposed in Ref.~\cite{Halpern2018}
which solves the time development of $\sqrt{\rho},\ \sqrt{\rho}u,\ \sqrt{p}$,
but the purpose is quite different;
they solve $\sqrt{p}$ to preserve the positivity of the pressure.
In this article, Eq.~(\ref{eq:2.4}) is discretized rather than Eq.~(\ref{eq:2.1}), with the central difference and implicit midpoint rule.
The discrete system is

\begin{align}
\left.\frac{\delta \mathbf{Q}}{\delta \mathbf{W}}\right|^{*}_i
\begin{bmatrix}
\dfrac{\sqrt{\rho}^{n+1}_i-\sqrt{\rho}^n_i}{\Delta t}\\
\\
\dfrac{\sqrt{\rho}u^{n+1}_i-\sqrt{\rho}u^n_i}{\Delta t}\\
\\
\dfrac{\sqrt{\rho}H^{n+1}_i-\sqrt{\rho}H^n_i}{\Delta t}
\end{bmatrix}
+
\left.\frac{\delta \mathbf{F}}{\delta \mathbf{W}}\right|^{*}_i
\begin{bmatrix}
\dfrac{\sqrt{\rho}^{*}_{i+1}-\sqrt{\rho}^{*}_{i-1}}{2\Delta x}\\
\\
\dfrac{\sqrt{\rho}u^{*}_{i+1}-\sqrt{\rho}u^{*}_{i-1}}{2\Delta x}\\
\\
\dfrac{\sqrt{\rho}H^{*}_{i+1}-\sqrt{\rho}H^{*}_{i-1}}{2\Delta x}
\end{bmatrix}
=
\begin{bmatrix}
0\\ \\ 0\\ \\ 0
\end{bmatrix},\label{eq:2.5}\\
\left.\frac{\delta \mathbf{Q}}{\delta \mathbf{W}}\right|^*_i=
\begin{bmatrix}
2\sqrt{\rho}^*_i & 0 & 0\\
\\
\sqrt{\rho}u^*_i & \sqrt{\rho}^*_i & 0\\
\\
\dfrac{1}{\gamma}\sqrt{\rho}H^*_i & \dfrac{\gamma-1}{\gamma}\sqrt{\rho}u^*_i & \dfrac{1}{\gamma}\sqrt{\rho}^*_i
\end{bmatrix},\ 
\left.\frac{\delta \mathbf{F}}{\delta \mathbf{W}}\right|^*_i=
\begin{bmatrix}
\sqrt{\rho}u^*_i & \sqrt{\rho}^*_i & 0\\
\\
\dfrac{\gamma-1}{\gamma}\sqrt{\rho}H^*_i & \dfrac{\gamma+1}{\gamma}\sqrt{\rho}u^*_i & \dfrac{\gamma-1}{\gamma}\sqrt{\rho}^*_i\\
\\
0 & \sqrt{\rho}H^*_i & \sqrt{\rho}u^*_i
\end{bmatrix},\label{eq:2.6}\\
f^{*}\equiv \frac{f^{n+1}+f^n}{2},\label{eq:2.7}
\end{align}
where $f$ is an arbitrary function.
Note that the spatial discretization can be performed by central differences with arbitrary order of accuracy
since the Jacobian matrices Eq.~(\ref{eq:2.6}) are spatially localized.
We can discuss the eigenstructure of this system by multiplying the inverse matrix of $\delta \mathbf{Q}/\delta \mathbf{W}$
to Eq.~(\ref{eq:2.5}) from left:

\begin{align}
\left(\left.\frac{\delta \mathbf{Q}}{\delta \mathbf{W}}\right|^{*}_i\right)^{-1}
\left.\frac{\delta \mathbf{F}}{\delta \mathbf{W}}\right|^{*}_i
=
\begin{bmatrix}
\dfrac{\hat{u}}{2} & \dfrac12 & 0\\
\dfrac{\gamma-1}{\gamma}\hat{H}-\dfrac{{\hat{u}}^2}{2} & \dfrac{\gamma+2}{2\gamma}\hat{u} & \dfrac{\gamma-1}{\gamma}\\
\dfrac{\gamma-1}{2}{\hat{u}}^3-\dfrac{2\gamma^2-3\gamma+2}{2\gamma}\hat{H}\hat{u} &
-\dfrac{\gamma^2+\gamma-2}{2\gamma}{\hat{u}}^2+\dfrac{2\gamma-1}{2}\hat{H} & \dfrac{2\gamma-1}{\gamma}\hat{u}
\end{bmatrix},\label{eq:2.8}
\end{align}
where $\hat{u}$ and $\hat{H}$ are defined as follows; they are very similar to the ``Roe average.''

\begin{align}
\hat{u}\equiv \frac{\sqrt{\rho}u^*_i}{\sqrt{\rho}^*_i}
=\frac{\sqrt{\rho}u^{n+1}_i+\sqrt{\rho}u^n_i}{\sqrt{\rho}^{n+1}_i+\sqrt{\rho}^n_i},\ 
\hat{H}\equiv \frac{\sqrt{\rho}H^*_i}{\sqrt{\rho}^*_i}
=\frac{\sqrt{\rho}H^{n+1}_i+\sqrt{\rho}H^n_i}{\sqrt{\rho}^{n+1}_i+\sqrt{\rho}^n_i}.\label{eq:2.9}
\end{align}

Every scheme has its own eigenstructure in discrete form.
The velocities of entropy and pressure waves are obtained as roots of a characteristic equation about the eigenstructure.
Generally speaking, roots of a cubic equation can be complex,
which clearly violates one of the most important regulations of the hyperbolic system.
The eigenstructure-preserving scheme consistently composes the Jacobian matrix with $\hat{u}$ and $\hat{H}$
even in discrete level, so the cubic equation is factorized to ensure the real eigenvalues.
A set of linearly independent eigenvectors and corresponding eigenvalues is as follows:

\begin{align}
\mathbf{k}_1=
\begin{bmatrix}
1 \\ \hat{u} \\ {\hat{u}}^2-\hat{H}
\end{bmatrix},\ 
\mathbf{k}_2=
\begin{bmatrix}
1 \\ \hat{u}-2\hat{c_\mathrm{s}} \\ \hat{H}+2\hat{c_\mathrm{s}}^2-2\hat{c_\mathrm{s}}\hat{u}
\end{bmatrix},\ 
\mathbf{k}_3=
\begin{bmatrix}
1 \\ \hat{u}+2\hat{c_\mathrm{s}} \\ \hat{H}+2\hat{c_\mathrm{s}}^2+2\hat{c_\mathrm{s}}\hat{u}
\end{bmatrix},\label{eq:2.10}\\
\lambda_1=\hat{u},\ 
\lambda_2=\hat{u}-\hat{c_\mathrm{s}},\ 
\lambda_3=\hat{u}+\hat{c_\mathrm{s}},\label{eq:2.11}
\end{align}
where $\hat{c_\mathrm{s}}=\sqrt{(\gamma-1)(\hat{H}-\hat{u}^2/2)}$ is the sound speed calculated by
$\hat{u}$ and $\hat{H}$.
Therefore, the hyperbolic structure is strictly maintained even in discrete level
unless the pressure becomes negative.

\section{Numerical experiment}\label{sec:3}
Here we discuss the eigenstructures calculated by the proposed scheme and a conventional method.
The conventional method discretizes Eq.~(\ref{eq:2.1}) directly rather than Eq.~(\ref{eq:2.4}),
with the central difference and implicit midpoint rule.
The flow field is initialized as follows:

\begin{align}
\rho(x)=1,\ u(x)=0.1\sin(2\pi x),\ p(x)=p_0,\ \gamma=\frac53 ,\ -0.5\le x\le 0.5,\label{eq:3.1}
\end{align}
where $p_0$ is the parameter in this study; it is dislayed in Table~\ref{tab:1}.
The maximum Mach number of case-B is greater than that of case-A in this experiment.
Here we employ $\Delta t=1/32$ and $\Delta x=1/32$, so the Courant number is much less than unity.
A periodic condition is used at the boundaries.
The simulations are performed for ten time-steps,
but the calculations of eigenvalues are performed only at the first step.
The discrete Jacobian matrices of the examined schemes are composed at every grid point
and the eigenvalue problems are algebraically solved with LAPACK.

\begin{table}[h]
\centering
\caption{\label{tab:1} Experimental conditions and brief results.}
\begin{tabular}{ccccc}\hline
Case & $p_0$ & Proposed & Conventional\\ \hline
A  & $10^{-3}$ & Unpolluted & Unpolluted\\
B  & $10^{-4}$ & Unpolluted & Polluted\\ \hline
\end{tabular}
\end{table}

\begin{figure}
\begin{minipage}{0.5\textwidth}
\centering
\includegraphics[width=\textwidth]{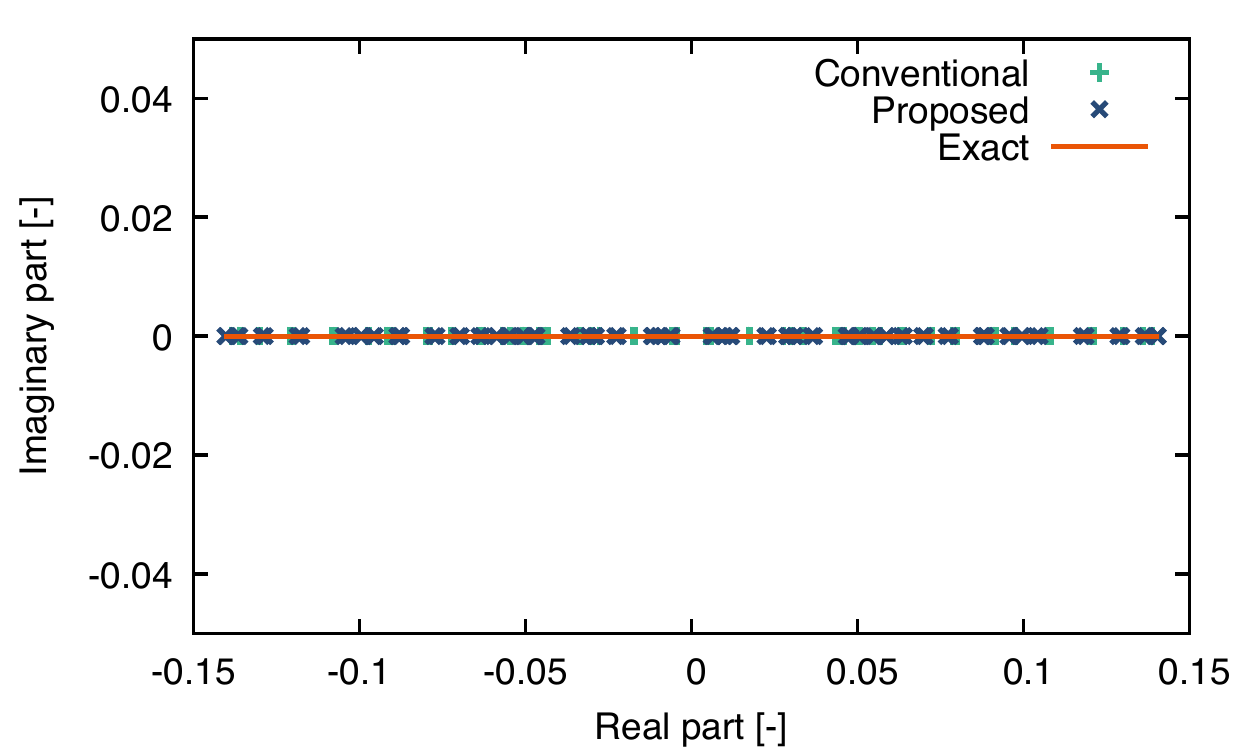}
\subcaption{\label{fig:1a} Case-A.}
\end{minipage}
\begin{minipage}{0.5\textwidth}
\centering
\includegraphics[width=\textwidth]{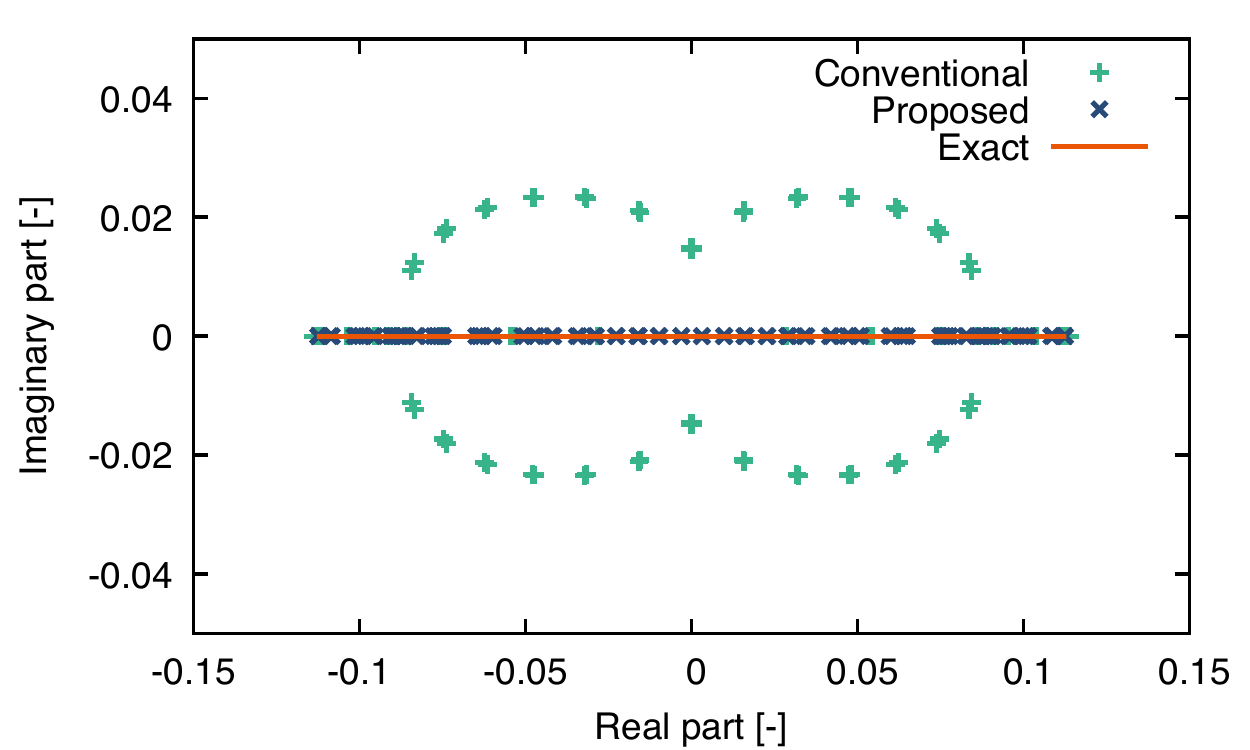}
\subcaption{\label{fig:1b} Case-B.}
\end{minipage}
\caption{\label{fig:1} (Color Online) Discrete eigenvalues in complex planes.}
\end{figure}

Figures~\ref{fig:1a} and \ref{fig:1b} show the discrete eigenvalues of case-A and case-B, respectively.
The discrete eigenvalues are also compared to the exact solution, i.e., $u,\ u\pm c_\mathrm{s}$.
In case-A whose maximum Mach number is $M \simeq 2.4$, both schemes reproduce the exact eigenvalues well.
In contrast, the conventional scheme generates complex eigenvalues in case-B ($M \simeq 7.7$),
which is not allowed in the hyperbolic system.
This means that the conventional scheme encounters a pollution of the eigenstructure.
The complex eigenvalues imply the conventional scheme includes a numerical dissipation in this situation.
If the coefficient of numerical dissipation is negative, this is an evidence of the numerical instability
since the negative dissipation amplifies numerical noises.
Unfortunately, we cannot demonstrate that this hypothesis is true only from this experiment.
What we can conclude is the eigenstructure-preserving scheme is free from such a pollution of the eigenstructure.

\begin{figure}
\begin{minipage}{0.5\textwidth}
\centering
\includegraphics[width=\textwidth]{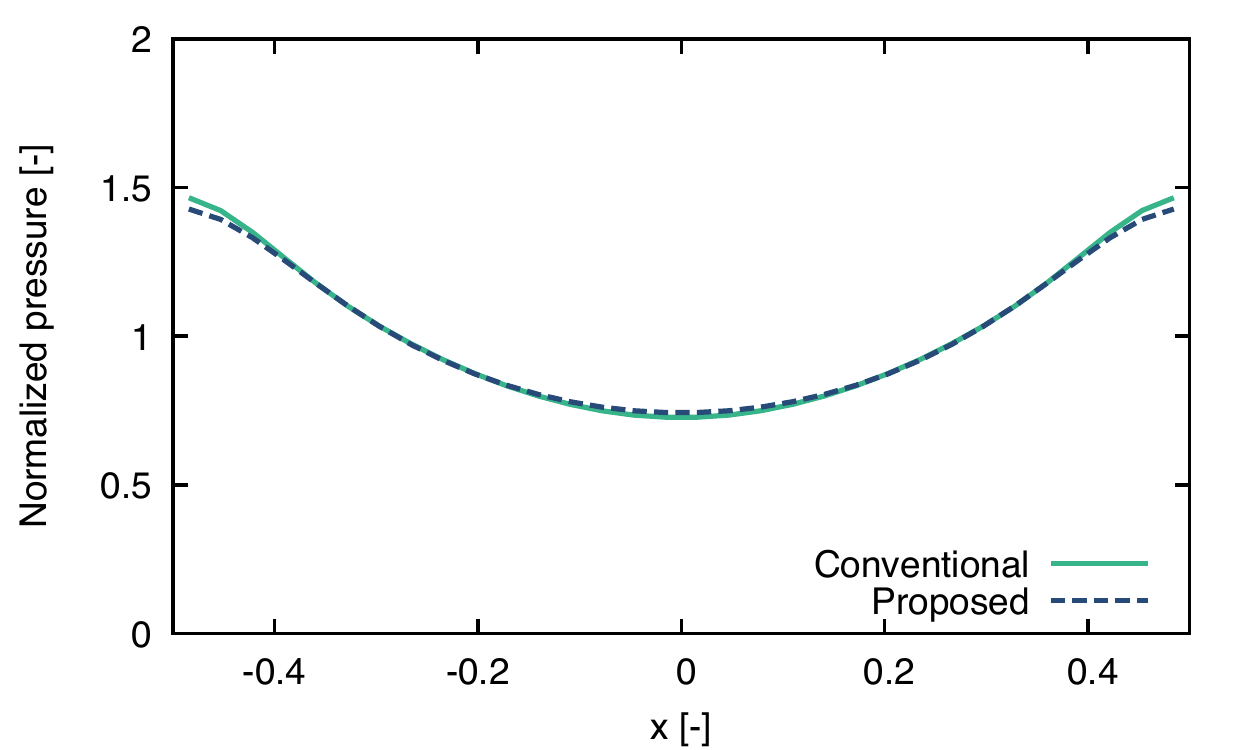}
\subcaption{\label{fig:2a} Case-A.}
\end{minipage}
\begin{minipage}{0.5\textwidth}
\centering
\includegraphics[width=\textwidth]{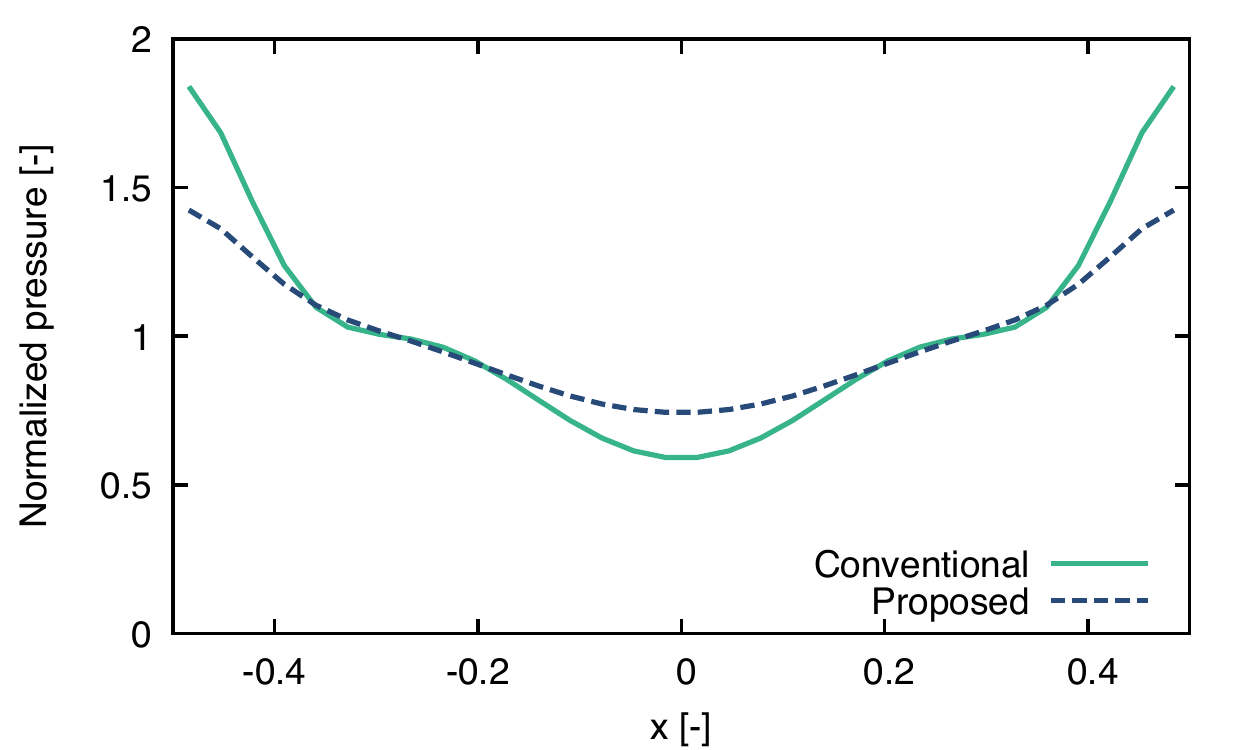}
\subcaption{\label{fig:2b} Case-B.}
\end{minipage}
\caption{\label{fig:1} (Color Online) Pressure profiles of the examined schemes.}
\end{figure}

\begin{figure}
\centering
\includegraphics[width=0.5\textwidth]{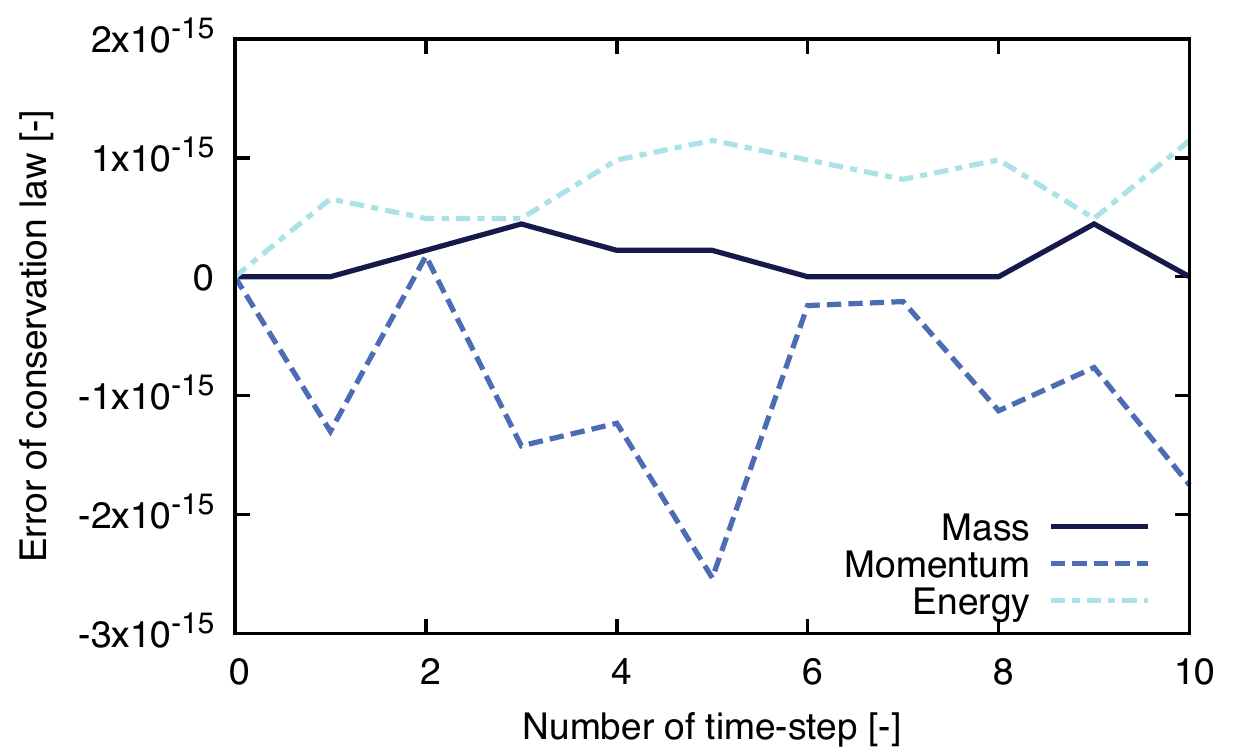}
\caption{\label{fig:3} (Color Online) Conservation properties of the proposed scheme (case-B).}
\end{figure}

Figures~\ref{fig:2a} and \ref{fig:2b} show the pressure profiles of the examined schemes in case-A and case-B, respectively.
The pressure at ten time-step normalized by the initial condition is displayed.
The conventional scheme has a numerical oscillation whose wavelength is shorter than that of the velocity perturbation
when the discrete eigenvalues are complex.
In contrast, the proposed scheme seems not to be affected by the unphysical mode.
Note that we focus on the local eigenstructure in this article,
so it is unclear why such a global numerical oscillation is suppressed by the proposed scheme.
Further, the conservation properties of the proposed scheme are described in Fig.~\ref{fig:3}.
The mass-momentum-energy conservation is strictly preserved at the round-off level of the double precision floating point number.

\section{Conclusions}\label{sec:4}
In this article, an eigenstructure-preserving scheme is proposed for the one-dimensional compressible Euler equation.
The scheme ensures the eigenvalues are always real (i.e., not complex) and the hyperbolic structure is strictly maintained in discrete level.
Although the eigenstructure is discussed in a non-conservative formulation,
the proposed scheme is locally conservative owing to the skew-symmetric operators.
It is demonstrated that the eigenstructure-preserving scheme generates real eigenvalues in a numerical experiment
with high Mach number where a conventional scheme possesses complex eigenvalues.
Further, the eigenstructure-preserving scheme is applicable not only to the finite-difference method but
also to the finite-element method, although the finite-volume version is difficult to be constructed.

\section*{Acknowledgment}
This work was carried out within the framework of Broader Approach.


\end{document}